\documentclass[10pt,conference]{IEEEtran}
\IEEEoverridecommandlockouts
\usepackage{cite}
\usepackage{amsmath,amssymb,amsfonts}
\usepackage{graphicx}
\usepackage{textcomp}

\usepackage{algorithm}
\usepackage{utfsym}
\usepackage{algpseudocode}
\usepackage{multirow}
\usepackage{booktabs}
\usepackage{makecell}
\usepackage{pifont}
\usepackage{tikz}
\usepackage{xifthen}
\usepackage[dvipsnames, svgnames, x11names]{xcolor}
\usepackage{subcaption}
\usepackage{hyperref}
\usepackage{doi}
\usepackage[firstpageonly=false]{draftwatermark}
\DraftwatermarkOptions{text={Preliminary Version},fontsize=0.08\paperwidth,color={[gray]{0.80}},angle=55}

\newcommand{\gou}{\textcolor{Green}{\ding{52}}}
\newcommand{\bangou}{\textcolor{ProcessBlue}{\ding{52}\rotatebox[origin=c]{-9.2}{\kern-0.7em\ding{55}}}}
\newcommand{\cha}{\textcolor{RedViolet}{\ding{55}}}

\definecolor{cG}{rgb}{0.35,0.75,0.5}
\definecolor{cR}{rgb}{0.75,0.1,0.1}
\def\BibTeX{{\rm B\kern-.05em{\sc i\kern-.025em b}\kern-.08em
    T\kern-.1667em\lower.7ex\hbox{E}\kern-.125emX}}
\begin{document}
\title{Rethinking Adaptive Routing on Dragonfly\thanks{This is a preliminary version of this work. An updated final version will follow, and the source code will be open-sourced.}}

\author{\IEEEauthorblockN{Ziyu Wang, Dezun Dong\textsuperscript{\textsection}, and Fei Lei}
\IEEEauthorblockA{College of Computer Science and Technology, National University of Defense Technology\\
\{wangziyu, dong, leifei\}@nudt.edu.cn}
\thanks{\textsuperscript{\textsection}Corresponding author: Dezun Dong (dong@nudt.edu.cn).}}

\maketitle

\begin{abstract}
The Dragonfly topology is widely deployed in leading supercomputers, where
adaptive routing is crucial for high-performance interconnection. Despite years
of research, a dilemma remains: state-of-the-art adaptive routing algorithms
achieve excellent performance but demand hardware budgets far beyond what real
router chips can afford---prohibitive buffers, logically dependent
floating-point operations, or exhaustive path-set traversals---whereas
realizable algorithms deliver unsatisfactory performance. Furthermore, existing
algorithms fail to exploit the abundant almost-minimal paths in Dragonfly: when
only the local channels of a minimal path are congested, they reroute packets
to significantly longer global non-minimal paths, wasting global bandwidth and
increasing latency. To address both limitations, we analyze intra-group path
contention and propose the \textbf{D}ecoupled \textbf{I}ncremental
\textbf{A}daptive \textbf{R}outing (DIAR) algorithm, which decouples routing
into global path selection at the source and incremental local detours onto
almost-minimal paths, guaranteeing deadlock freedom without additional virtual
channels, escape rings, or turn models. Requiring only a few comparators,
shifters, adders, and multiplexers, DIAR reduces buffer overhead by up to four orders of magnitude
compared with state-of-the-art algorithms, while achieving comparable or
superior performance across all evaluated scenarios: relative to the
strongest realizable adaptive routing algorithm in each scenario, DIAR
improves saturation throughput by up to 30.0\% across synthetic traffic
patterns and improves performance by up to 34.9\% (13.2\% on average)
across HPC workloads.
\end{abstract}


\section{INTRODUCTION}
High-radix topologies such as the Flattened Butterfly \cite{Flattend_butterfly07}, HyperX \cite{HyperX09}, and Dragonfly \cite{Dragonfly08} offer favorable cost-effectiveness and low network diameters, and have been deployed in large-scale interconnection systems. As of June 2026 \cite{Top500_202606}, six of the ten most powerful HPC clusters \cite{frontier22, elcapitan25, aurora23, hpc7sc, hpc6sc, Alps} use a Dragonfly-based topology (the Slingshot-11 interconnect \cite{slingshot20}).

In Dragonfly networks, the global channels connecting any two groups are highly sparse, making adaptive routing essential: when the global channels along the minimal path are congested, traffic is diverted to non-minimal paths (i.e., forwarded via an intermediate group) to achieve load balancing across global channels. Universal Globally-Adaptive Load-balanced (UGAL \cite{ugal05}) routing has been deployed in practical Dragonfly systems \cite{par2w15, DGB22}. Since the original UGAL delivers unsatisfactory performance, a variety of UGAL-based adaptive routing algorithms have been proposed. However, these algorithms commonly suffer from the following two limitations.

\begin{figure}[t]
  \centering
  \includegraphics[width=0.99\linewidth]{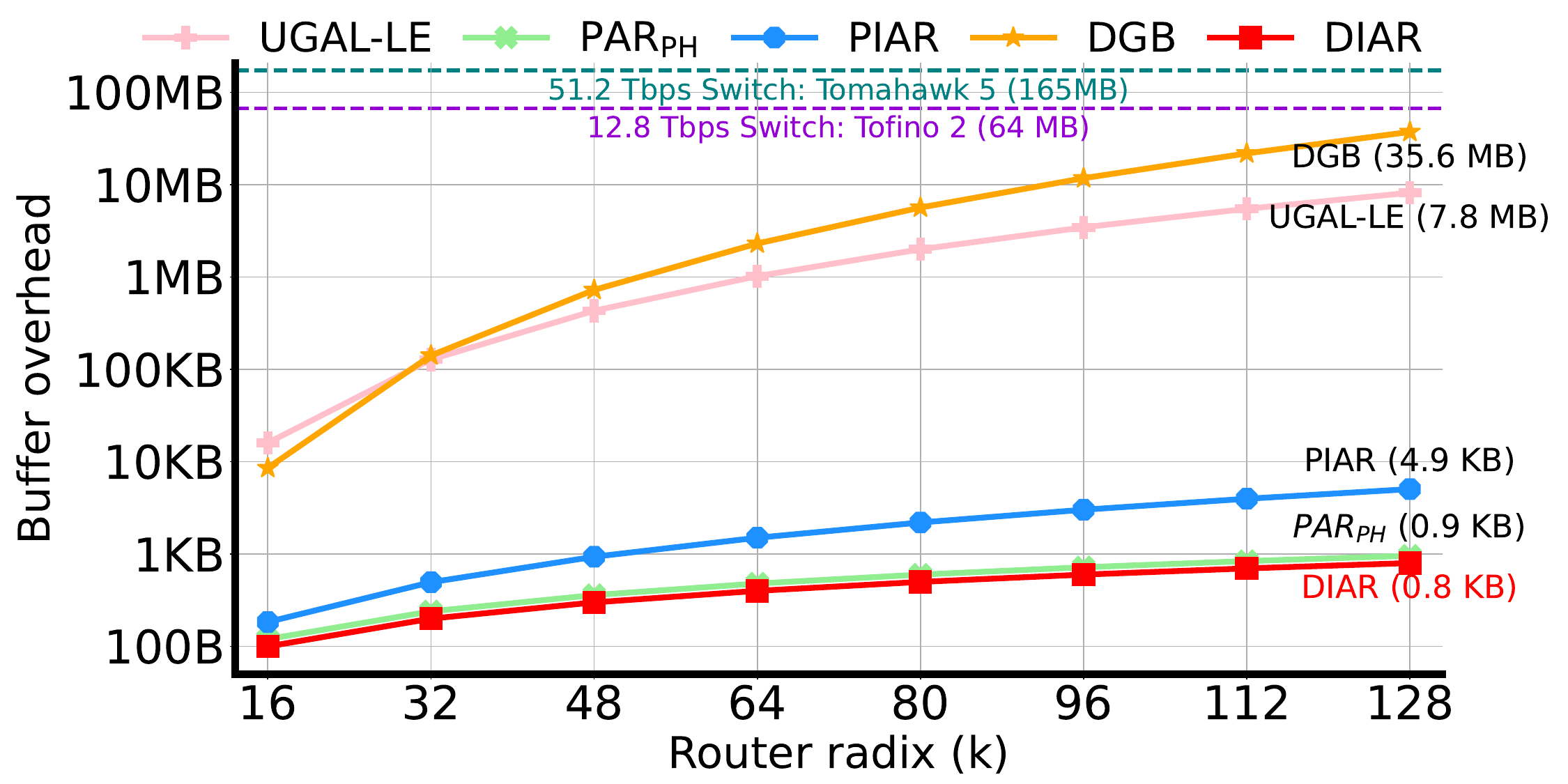}
  \caption{Buffer overhead comparison of various advanced adaptive routing algorithms across different router radixes.}
  \label{buffer_cost}
\end{figure}

\textbf{Limitation 1: failing to strike a good balance between performance and realizability.} The primary design goals of router/switch chips are higher throughput and lower forwarding latency, and adaptive routing is in fact a scarce, even luxurious, feature: the vast majority of Ethernet switches forward packets exclusively along shortest paths via ECMP hash-based table lookups, offering no support for non-minimal or adaptive routing; only InfiniBand and HPC-customized router chips (e.g., Cray Aries \cite{cray12, crayRout17} and Slingshot Rosetta \cite{slingshot20}) support this feature. Even in these chips, the implementation budget for adaptive routing is extremely limited---introducing complex floating-point computations or traversing an entire path set on the routing critical path is out of the question. Measured against this yardstick, DGB \cite{DGB22} and PIAR \cite{piar25}, the best-performing adaptive routing algorithms on Dragonfly to date, fall short: as shown in Figure~\ref{buffer_cost}, DGB incurs prohibitive storage overhead and its routing decisions require a chain of logically dependent floating-point multiplications and divisions, while PIAR, despite its lower storage overhead, must sequentially traverse hundreds to thousands of candidate paths in the path sets for every routing decision. These requirements render both nearly impossible to implement in real router chips. On the other hand, $\rm{PAR_{PH}}$ comes with acceptable buffer and logic overheads, but delivers unsatisfactory overall performance. The hardware overheads of these algorithms are further discussed in Section~\ref{Overhead}.

\begin{figure}[t]
  \centering
  \includegraphics[width=0.98\linewidth]{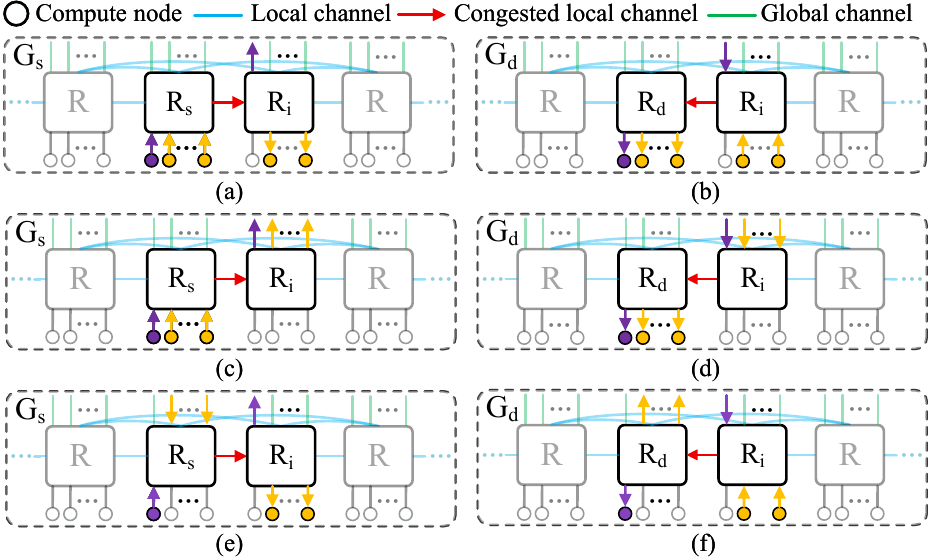}
  \caption{Six cases of local channel congestion. The purple and yellow nodes belong to distinct jobs or belong to the same job but are in different communication phases. (a) and (b): the intra-group traffic generated by the yellow nodes causes local channel congestion, thereby interfering with the inter-group communication of the purple nodes. (c), (d), (e), and (f): the minimal paths of the purple and yellow nodes traverse distinct global channels but share the same local channel.}
  \label{LocalCongestion}
\end{figure}

\textbf{Limitation 2: failing to exploit the abundant almost-minimal paths in Dragonfly.} Near-neighbor communication is highly prevalent in HPC workloads, and the resulting intra-group traffic can interfere with the inter-group communication of other nodes. As illustrated in Figure~\ref{LocalCongestion}(a), the intra-group traffic generated by the yellow nodes congests the local channel between the source router and router $R_i$. Similarly, as shown in Figure~\ref{LocalCongestion}(b), intra-group traffic also congests the local channel between router $R_i$ and the destination router $R_t$. Furthermore, inter-group communication can likewise cause local channel congestion, as shown in Figure~\ref{LocalCongestion}(c), (d), (e), and (f). In all six scenarios above, the bottleneck of the minimal path is a local channel rather than a global channel; in Section~\ref{congestion_model}, we will demonstrate that these scenarios are prevalent under common traffic patterns. However, when congestion backpressure reaches the source router, UGAL and other source-adaptive routing algorithms cannot tell whether the congestion originates from a local channel or a remote global channel, and thus conservatively resort to global non-minimal routing, detouring traffic to a random intermediate group---doubling global channel resource consumption.

\begin{figure}[t]
  \centering
  \includegraphics[width=0.98\linewidth]{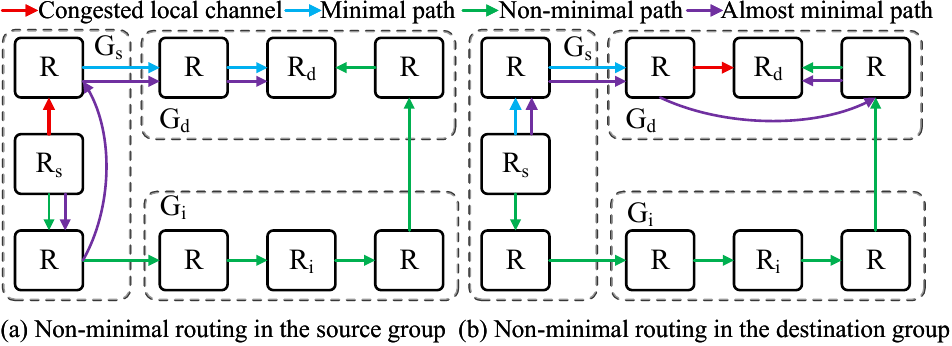}
  \caption{Two types of almost-minimal paths in Dragonfly.}
  \label{almostPath}
\end{figure}

In fact, a far cheaper alternative exists: as shown in Figure~\ref{almostPath}, a packet only needs to detour to a random router within the source and/or destination group, bypassing the bottleneck local channel at the cost of merely one extra local hop. Inspired by \cite{fatpath20}, we refer to such a path---one that uses a global channel only once (the same as minimal paths) but involves intra-group non-minimal routing---as an \textbf{almost-minimal path}. Compared with conventional non-minimal paths, which introduce an additional global hop and two additional local hops, almost-minimal paths substantially reduce global bandwidth consumption and network latency. Unfortunately, the one-shot decision paradigm of UGAL-style source routing is inherently incapable of exploiting such paths, while prior incremental adaptive routing algorithms that permit local detours \cite{PAR09, on_the_fly12} either prohibit them entirely or impose restrictive conditions that limit their effectiveness. Worse still, these approaches enlarge the maximum path hop count and thereby exacerbate the risk of deadlock; existing deadlock-avoidance solutions---additional virtual channels (VCs) \cite{PAR09}, Hamiltonian escape rings \cite{ofarcm13}, or turn models \cite{efficient2013}---all incur substantial buffer overhead or performance penalties.

Therefore, an ideal adaptive routing algorithm for Dragonfly should match or even surpass the state-of-the-art performance within a hardware budget comparable to that of Cray Aries---precisely the goal of this paper. To this end, we propose \textbf{D}ecoupled \textbf{I}ncremental \textbf{A}daptive \textbf{R}outing (DIAR). DIAR decouples routing decisions into global path selection at the source and incremental local non-minimal routing, governed by a set of Restrictive Incremental Adaptive Routing (RIAR) rules: packets designated for global minimal paths may take local detours in the source and/or destination groups when beneficial, while global non-minimal packets maintain their traditional paths. This design fully exploits almost-minimal paths for load balancing without increasing the longest path length, thereby avoiding deadlock without any extra hardware resources or path restrictions. Moreover, DIAR requires only lightweight buffer/logic overhead and is fully compatible with current router architectures. The contributions of this work include the following.

\begin{itemize}
    \item We define almost-minimal paths in Dragonfly and demonstrate their significance via a path contention model.
    \item We propose the DIAR algorithm, which fully exploits almost-minimal paths while guaranteeing deadlock freedom without additional VCs, escape rings, or turn models.
    \item We thoroughly analyze the implementation complexity of various adaptive routing algorithms: compared with state-of-the-art algorithms, DIAR reduces the buffer overhead by up to four orders of magnitude and involves no floating-point operations or path-set traversals.
    \item We conduct extensive evaluations. Compared with state-of-the-art routing algorithms on Dragonfly, DIAR consistently achieves comparable or superior performance.
    
\end{itemize}

\section{BACKGROUND}
\subsection{Dragonfly Topology}
\begin{figure}[t]
  \centering
  \includegraphics[width=0.92\linewidth]{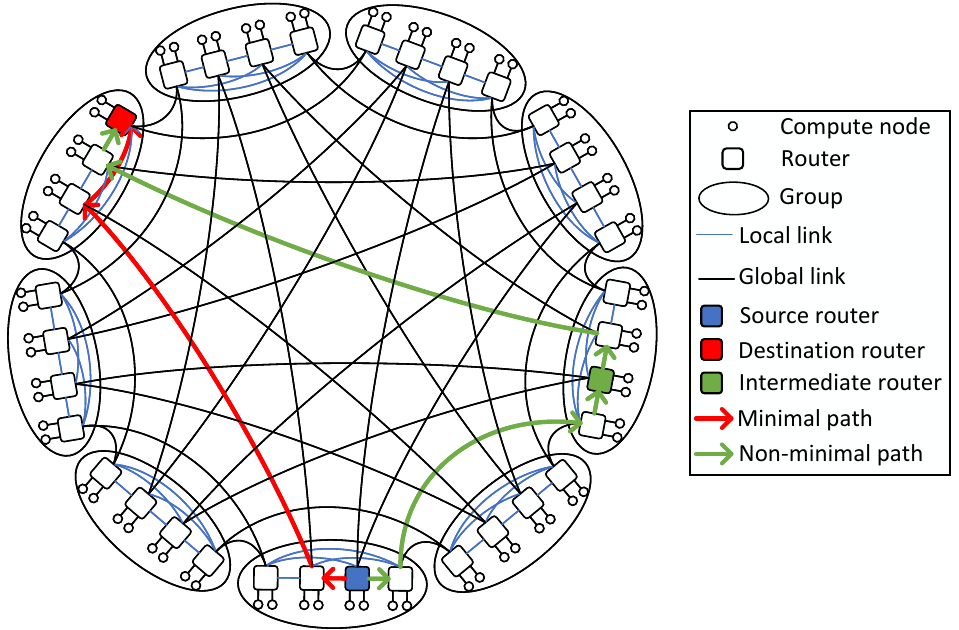}
  \caption{The minimal and non-minimal paths in $dfly(2,4,2,9)$.}
  \label{dragonfly_topo}
\end{figure}

As illustrated in Figure \ref{dragonfly_topo}, a Dragonfly network consists of multiple fully interconnected groups, with each group comprising a set of routers that are also fully interconnected. A Dragonfly network is uniquely defined by four key parameters: \( p \), \( a \), \( h \), and \( g \). We can use the notation $dfly(p,a,h,g)$ to represent any Dragonfly topology. Each router has $p$ channels to connect to nodes, $a-1$ local channels to connect to other intra-group routers, and $h$ global channels to connect to routers in other groups. For load-balanced Dragonfly networks, the parameters satisfy \(a = 2p = 2h\) \cite{Dragonfly08}. \(g\) represents the number of groups in the network. Given the parameters \(a\) and \(h\), a Dragonfly network can support up to \(ah+1\) groups, where only one global channel connects any two groups.

In Figure~\ref{dragonfly_topo}, the red path represents the minimal path, where the packet departs from the source router, uses the global channel only once, and reaches the destination router within at most three hops. The green path represents the non-minimal path (i.e., Valiant path \cite{val82}), in which the packet is first minimally routed to an intermediate router randomly selected from the network, and then minimally routed again to the destination router. The green path in Figure~\ref{dragonfly_topo} is the Valiant-node (VALn) path; the original Valiant-group variant \cite{Dragonfly08} saves one local hop but may congest local channels in the intermediate group \cite{on_the_fly12}, so subsequent algorithms \cite{par2w15, ECtN15, DGB22, TPR18, ugal-le24}, including those evaluated here, default to VALn.

The intra-group topology of Dragonfly can also adopt a 2D flattened butterfly \cite{cray12} or a two-level fat-tree structure \cite{dragonfly+17, megafly18, jupiter25}. Considering that leadership-class supercomputers such as El Capitan \cite{elcapitan25}, Frontier \cite{frontier22}, and Aurora \cite{aurora23} employ fully connected intra-group Dragonfly networks, this paper, like most prior studies, focuses on Dragonfly networks with fully connected intra-group topology.

\subsection{Adaptive Routing on Dragonfly}

UGAL \cite{ugal05} is the default source adaptive routing algorithm for Dragonfly and has been deployed in large-scale interconnection systems \cite{cray12}. UGAL estimates congestion for the minimal and Valiant \cite{val82} paths by the product of the next-hop input queue occupancy and the path hop count, and then determines the packet routing strategy. Specifically, UGAL chooses the minimal path when \eqref{ugal_eq} is satisfied:
\begin{equation} \label{ugal_eq}
    q_m \times H_m \leq q_{nm} \times H_{nm} + b
\end{equation}
where \(q_m\) (\(q_{nm}\)) denotes the queue occupancy along the minimal (non-minimal) path. \(H_m\) (\(H_{nm}\)) represents the hop count of the minimal (non-minimal) path. \(b\) is a static bias term: a positive value of \(b\) biases UGAL toward selecting the minimal path, whereas a negative value biases it toward the non-minimal path.

A variety of adaptive routing algorithms have been proposed to improve the performance of UGAL. Based on their core motivations, these algorithms can be classified into three categories, and their current limitations are analyzed respectively:

\textbf{1) Improving path congestion awareness: }
To address the indirect routing issue \cite{Dragonfly08} in Dragonfly---where source routers perceive remote global channel congestion solely via delayed and inaccurate backpressure---PiggyBack (PB \cite{PAR09}) routing was proposed. Specifically, a global channel is identified as congested when \eqref{pb_eq} holds, where $q$ and $q_{avg}$ denote its queue length and the average queue length of other global
\begin{equation} \label{pb_eq}
    q-2 \times q_{avg} > T
\end{equation}
channels, respectively, and $T$ is the congestion threshold. Each router then piggybacks the congestion bitmap of its connected global channels to other intra-group routers. To mitigate phantom congestion \cite{par2w15}---where many in-flight flits and credits on a normally flowing, long-distance channel cause high queue occupancy and false congestion sensing---$\rm{PAR_{PH}}$ \cite{par2w15} pioneered a sliding window technique. Specifically, it replaces $q_m$ and $q_{nm}$ in \eqref{ugal_eq} with $q_m'$ and $q_{nm}'$ from \eqref{par2w}, where $L$ is the channel latency and $W(2L)$ records
\begin{equation} \label{par2w}
    q'=q-W(2L)
\end{equation}
the number of flits transmitted over the channel within the recent round-trip time. This sliding-window traffic tracking technique was subsequently adopted by many algorithms \cite{acor_pro25, TPR18, ugal-le24, DGB22, piar25}. TPR \cite{TPR18} uses local sliding windows to infer network traffic patterns and adjust UGAL's bias. Q-adaptive \cite{Q_ada21} and DGB \cite{DGB22} employ multi-agent reinforcement learning and gradient descent, respectively, to learn global congestion from end-to-end latency, achieving high performance across diverse traffic patterns. UGAL-LE \cite{ugal-le24} observes non-minimal traffic from the source group via local counters to infer global channel congestion. 
\textbf{Limitations:} Intelligent algorithms like Q-adaptive and DGB achieve high performance but incur significant buffer and logic overheads, whereas non-intelligent algorithms offer simpler implementations but limited performance gains.

\textbf{2) Optimizing global non-minimal path hop counts: }While the majority of global non-minimal paths in the Dragonfly network span 6 hops, shorter global non-minimal paths of 2, 3, 4, and 5 hops also exist \cite{piar25}. T-UGAL \cite{T_ugal19} employs a customized set of shorter global non-minimal paths, while ACOR \cite{ACOR19} and LIA \cite{acor_pro25} select global non-minimal paths with varying hop counts based on local traffic states and counters, respectively. PIAR \cite{piar25} partitions multiple global non-minimal path sets according to path hop counts; during routing decisions, it traverses all paths within these sets and routes packets to the path with the least congestion and the fewest possible hops. Furthermore, when evaluating path congestion, PIAR combines \eqref{pb_eq} and \eqref{par2w} to propose \eqref{piar_eq}. Specifically, when \eqref{piar_eq} is satisfied,
\begin{equation} \label{piar_eq}
    q-W(2L) > T
\end{equation}
the channel is identified as congested. Subsequently, similar to the PB algorithm, the router also piggybacks the congestion bitmap of the global channels. \textbf{Limitations: } The distribution of shorter global non-minimal paths between each source-destination router pair is highly variable and unpredictable. Therefore, routers must either store these shorter non-minimal paths---incurring substantial buffer overhead---or perform additional computation to identify them dynamically.
Moreover, current high-radix routers implement adaptive routing by randomly sampling only a handful of paths (e.g., two minimal and two non-minimal) due to hardware limitations \cite{cray12}. In contrast, PIAR necessitates traversing hundreds of candidate paths within each path set, requiring substantial architectural modifications that undermine its deployability \cite{piar25}.

\textbf{3) Enabling incremental adaptive routing: }Addressing the indirect routing issue, PAR \cite{PAR09} allows minimally routed packets an additional opportunity to switch to non-minimal routing before leaving the source group. Furthermore, OFAR \cite{on_the_fly12} enables an intra-group rerouting opportunity within any traversed group. 
\textbf{Limitations: }Such incremental adaptive routing algorithms avoid deadlocks at the cost of additional VCs or reduced path diversity, which will be discussed in detail in Section \ref{incremental_limitation}.

The DIAR algorithm proposed in this paper also falls into the third category. Local misrouting itself is not new \cite{on_the_fly12, efficient2013}, but no prior work exploits intra-group detours for packets already committed to a global minimal path: DIAR is the first to define the almost-minimal paths of Dragonfly and to exploit all three of them within the standard four-VC budget, while maintaining good hardware affinity.

\section{MOTIVATION}

\subsection{Local Channel Congestion}

\label{congestion_model}

Figure~\ref{LocalCongestion} demonstrates that there exist multiple scenarios where the local channels of the minimal path are congested while the global channels remain uncongested. What proportion do these scenarios account for under common network traffic patterns? To clarify this, we need a finer-grained classification of the congestion states along the minimal path. A minimal path in a Dragonfly network utilizes at most three channels: a local channel within the source group, a global channel, and a local channel within the destination group. Since each channel exhibits a binary state (either congested or uncongested), there are a total of $2^3 = 8$ distinct congestion scenarios for the minimal path. The distribution of these congestion scenarios depends heavily on the traffic pattern. Under Uniform Random (UR) traffic, load is naturally balanced and congestion rarely occurs, whereas Adversarial (ADV) traffic severely oversubscribes the global channels between adjacent groups. We therefore focus on Random Permutation (PERM)---a widely investigated pattern \cite{fatpath20, ics20_global_connect, REPS26, Uno25, DGB22, zettafly25, ugal-le24, par2w15} where each node communicates with exactly one random destination, representing irregular workloads that more closely resemble real-world scenarios \cite{fatpath20, ics20_global_connect}.

We conducted a static analysis of minimal-path contention that occurs during communication between all sender--receiver pairs, on four Dragonfly networks of increasing scale (1,056 to 40,200 nodes), averaged over 100 random permutations. Specifically, we assign an index to each channel in the network; a channel is contended when it is shared by two or more minimal paths, and each minimal path is classified by the binary congestion states of its source-group local, global, and destination-group local channels (e.g., ``000'' denotes a minimal path without any contention). The fraction of paths without any contention decreases with increasing network size, indicating that larger networks are more prone to path contention, whereas the fraction of paths experiencing contention on all three channels (the most severe case) increases as the network scales up.

The global non-minimal path can resolve all types of channel congestion, as it is implemented by randomly selecting an intermediate router across the entire network. Essentially, it serves as a universally effective path that achieves full load balancing. However, this comes at the cost of six hops and additional network resource consumption. Therefore, global non-minimal routing should be employed selectively, only when global channel congestion occurs (path types ``010'', ``011'', ``110'', ``111'').
In contrast, path types ``001'', ``100'', and ``101'' exhibit no global channel congestion; instead, they experience congestion solely in the destination group local channels, the source group local channels, and both local channel domains, respectively. Notably, across the four varying-sized Dragonfly networks, these three types consistently constitute approximately one-third of all contended paths, indicating that local-channel contention represents a substantial fraction of scenarios that cannot be overlooked.
 
Traffic associated with these three congestion types should utilize almost-minimal paths for load balancing; however, all prior adaptive routing algorithms either fail to exploit or underutilize these paths, thereby wasting substantial global channel bandwidth and resulting in high network latency.

\subsection{Limitations of Previous Incremental Adaptive Routings}
\label{incremental_limitation}

There are two primary incremental adaptive routing strategies in Dragonfly networks: PAR \cite{PAR09} and OFAR \cite{on_the_fly12}, which differ in the number of rerouting operations. PAR only allows packets determined by the source router to follow the global minimal route to be rerouted to a global non-minimal route when they arrive at a router within the source group that has a direct connection to the destination group. In contrast, OFAR additionally permits packets to adopt an intra-group non-minimal routing upon reaching the destination group, based on channel congestion. Specifically, these two incremental adaptive routing algorithms have the following two limitations:

\textbf{1) Compared with UGAL, extending the upper bound of the maximum hop count requires either additional VCs or the introduction of Hamiltonian escape rings or turn models to ensure deadlock-freedom.} High-radix networks typically adopt the deadlock-free strategy proposed in \cite{VCDeadlock81}, which allocates a number of VCs equal to the hop length of the network's longest path. Dragonfly is a diameter-3 topology whose longest minimal path has at most three hops. However, this minimal path requires only two VCs to ensure deadlock-freedom. This is because minimal paths are restricted to using at most one global channel; consequently, in cycles formed exclusively by minimal paths, no minimal path traverses two global channels. Therefore, global channels can reuse the VCs allocated to local channels. A non-minimal path can be defined as the concatenation of two consecutive minimal paths and requires a total of four VCs. Let $L$ and $G$ represent local hops and global hops, respectively. The VC usage sequence for a minimal path is $L_{vc2} \rightarrow G_{vc2} \rightarrow L_{vc3}$, and that for a global non-minimal path is:
\begin{equation} \label{non_min6}
    L_{vc0} \rightarrow G_{vc0} \rightarrow L_{vc1} \rightarrow L_{vc2} \rightarrow G_{vc2} \rightarrow L_{vc3}
\end{equation}

PAR \cite{PAR09} routing allows a packet that initially follows a minimal route to switch to a non-minimal route before departing the source group; its longest path \eqref{non_min7} requires five VCs. OFAR further permits packets to perform one intra-group non-minimal route in the destination group; without relying on Hamiltonian escape rings or turn models, its longest path \eqref{non_min8} requires six VCs:
\begin{align}
&\resizebox{0.74\linewidth}{!}{$L_{vc0} \rightarrow L_{vc1} \rightarrow G_{vc1} \rightarrow L_{vc2} \rightarrow L_{vc3} \rightarrow G_{vc3} \rightarrow L_{vc4}$} \label{non_min7}\\
&\resizebox{0.85\linewidth}{!}{$L_{vc0}\rightarrow L_{vc1}\rightarrow G_{vc1} \rightarrow L_{vc2} \rightarrow L_{vc3} \rightarrow G_{vc3} \rightarrow L_{vc4} \rightarrow L_{vc5}$} \label{non_min8}
\end{align}

In large-scale off-chip networks, VC resources are highly valuable for covering the round-trip latency of long links---practical systems such as the Cray XC \cite{cray12} implement only four routing VCs \cite{VVQ2023}; meanwhile, Hamiltonian escape rings suffer from limited bandwidth and long paths, and turn models sacrifice path diversity. Implementing incremental adaptive routing in Dragonfly is thus highly challenging.

\textbf{2) Insufficient exploitation of almost-minimal paths:} Both PAR and OFAR rely solely on backpressure to detect remote congestion. When the source router perceives congestion on the minimal path, it cannot determine whether the congestion is due to a local channel or caused by global channel backpressure. Consequently, both algorithms conservatively resort to global non-minimal routing, forgoing opportunities to utilize alternative paths with lower overhead.
Specifically, neither PAR nor OFAR can leverage the almost-minimal paths shown in Figure~\ref{almostPath}(a) for traffic with congestion types ``100'' and ``101''. Their capabilities diverge for other scenarios: OFAR permits packets to route non-minimally within the destination group, thereby enabling utilization of the almost-minimal paths in Figure~\ref{almostPath}(b). Even there, the detour is a by-product of its per-hop backpressure trigger rather than a path class to be exploited: OFAR's local misrouting exists to relieve the local saturation created by globally non-minimal traffic~\cite{on_the_fly12, efficient2013}. PAR, in turn, prohibits intra-group non-minimal routing in the destination group and thus cannot exploit any almost-minimal
paths. Table~\ref{increment_table} summarizes the mechanisms of all known adaptive routing algorithms for Dragonfly networks (ordered by publication date), among which only DIAR can fully exploit all almost-minimal paths and achieve deadlock avoidance without requiring additional resources.

\begin{table}[t]
\footnotesize
  \caption{Comparison of routing mechanisms of various adaptive routing algorithms on Dragonfly. Symbols \gou, \bangou, and \cha ~ denote ``support,'' ``partially support,'' and ``not support,'' respectively. ``Simple deadlock-free mechanism'' refers to a method that, compared with UGAL, does not require additional VCs, escape rings, or turn models to prevent deadlock. Q-adaptive and PIAR additionally grant packets an opportunity to select VALg or VALn routing within intermediate groups. Although not introducing new incremental routing mechanisms, we still classify them as incremental adaptive routing.}
  \label{increment_table}
  \setlength{\tabcolsep}{3pt}
  \begin{tabular}{|c|c|c|c|}
    \hline
    Algorithm&\makecell{Incremental\\adaptive routing}&\makecell{Utilizing almost-\\minimal paths}&\makecell{Simple deadlock-\\free mechanism}\\
    \hline
    UGAL\cite{ugal05} & \cha  & \cha & \gou \\
    \hline
    PB\cite{PAR09} & \cha & \cha  & \gou \\
    \hline
    PAR\cite{PAR09} & \gou & \cha  & \cha \\
    \hline
    OFAR\cite{on_the_fly12, efficient2013} & \gou & \bangou  & \cha  \\
    \hline
    ECtN \cite{ECtN15} & \cha & \cha & \gou  \\
    \hline
    $\rm{PAR_{PH}}$ \cite{par2w15} & \gou & \cha & \cha  \\
    \hline
    TPR \cite{TPR18} & \cha & \cha & \gou \\
    \hline
    ACOR \cite{ACOR19} & \cha & \cha & \gou  \\
    \hline
    T-UGAL \cite{T_ugal19} & \cha & \cha & \gou  \\
    \hline
    Q-adaptive \cite{Q_ada21} & \gou & \cha & \gou \\
    \hline
    DGB \cite{DGB22} & \cha & \cha & \gou \\
    \hline
    UGAL-LE \cite{ugal-le24} & \cha & \cha & \gou \\
    \hline
    LIA \cite{acor_pro25} & \cha & \cha & \gou \\
    \hline
    PIAR \cite{piar25} & \gou & \cha & \gou \\
    \hline
    DIAR (Ours) & \gou & \gou & \gou \\
    \hline
\end{tabular}
\end{table}

\section{DIAR ROUTING DESIGN}
In this section, we introduce the Decoupled Incremental Adaptive Routing (DIAR) algorithm. DIAR supports incremental adaptive routing and effectively exploits all types of almost-minimal paths in the network, while requiring no additional virtual channels, escape rings, or turn models for deadlock-freedom.

\subsection{Restrictive Incremental Adaptive Routing}

This section introduces the proposed Restrictive Incremental Adaptive Routing (RIAR) mechanism---the core of DIAR. RIAR operates under the following three routing rules:  
(1) Packets are allowed to perform incremental local non-minimal routing.  
(2) The global adaptive routing strategy must be determined at the source router. 
(3) Only packets identified as global minimal routes at the source router are allowed to perform local non-minimal routing within the source group and the destination group.  

The core design principle of RIAR is to fully exploit almost-minimal paths while strictly bounding the maximum hop count to six hops---matching the longest default non-minimal path in Dragonfly (at most four local hops and two global hops). As defined before, Rules (2) and (3) distinguish RIAR from PAR and OFAR by preventing path sequences exceeding six hops (illustrated in (\ref{non_min7}) and (\ref{non_min8})). Consequently, RIAR provably avoids deadlock with only four VCs while adopting the default global non-minimal path. Additionally, under Rules (2) and (3), RIAR leverages all three types of almost-minimal paths:

\begin{equation} \label{am_seqs}
\begin{aligned}
\text{``100'': }\; & L_{vc0} \rightarrow L_{vc1} \rightarrow G_{vc1} \rightarrow L_{vc3}\\
\text{``001'': }\; & L_{vc1} \rightarrow G_{vc1} \rightarrow L_{vc2} \rightarrow L_{vc3}\\
\text{``101'': }\; & L_{vc0} \rightarrow L_{vc1} \rightarrow G_{vc1} \rightarrow L_{vc2} \rightarrow L_{vc3}
\end{aligned}
\end{equation}

which resolve the ``100'', ``001'', and ``101'' types of path congestion, respectively. All almost-minimal paths employ exactly one global hop and at most four local hops.

\textbf{Deadlock freedom.} Order the channel classes as $L_{vc0} \prec G_{vc0} \prec L_{vc1} \prec G_{vc1} \prec L_{vc2} \prec G_{vc2} \prec L_{vc3}$, where $L_{vci}$ ($G_{vci}$) denotes a local (global) channel on VC $i$. Under RIAR, global non-minimal packets keep the classical assignment~\eqref{non_min6}, whereas every global minimal packet uses $L_{vc0}$ for the source-group detour hop (if taken), $L_{vc1}$ to exit the source group, $G_{vc1}$ for the global hop, $L_{vc2}$ for the destination-group detour hop (if taken), and $L_{vc3}$ for the final hop---yielding the three sequences in~\eqref{am_seqs} and, when no detour is taken, $L_{vc1}{\rightarrow}G_{vc1}{\rightarrow}L_{vc3}$. Hence every path type acquires channels in strictly ascending order of $\prec$; the channel dependency graph is therefore acyclic, and RIAR is deadlock-free with four VCs~\cite{VCDeadlock81}. Note that global minimal packets enter the global channel on $G_{vc1}$ rather than $G_{vc2}$ precisely to reserve ascending VCs for a possible destination-group detour, whose necessity is unknown before arrival; after $G_{vc1}$, both the detour ($L_{vc2}$) and direct ($L_{vc3}$) continuations ascend, so the decision made upon arrival is always safe. All remaining routing choices in the implementation only truncate paths or skip forward in the VC order, so the argument covers every branch.

\subsection{Global Channel Congestion-Aware Design}

\label{sec42}

The challenge in implementing the RIAR mechanism lies in Rule (2), which requires the source router to perceive congestion conditions on both minimal and non-minimal global channels, and to make accurate global adaptive routing decisions. However, in Dragonfly networks, the minimal and non-minimal global channels are often not directly connected to the source router; instead, they reside on other routers within the same group \cite{PAR09, ugal-le24}. 

A straightforward approach, as used in UGAL, is to indirectly sense global channel congestion through backpressure. Nevertheless, this method is neither timely nor accurate. As shown in Figure~\ref{LocalCongestion}(c)--(f), minimal paths utilizing different global channels may share the same local channel. In such cases, the source router cannot determine whether the congestion is due to a local channel or a remote global channel and thus cannot distinguish between path congestion types ``10X'' and ``01X,'' where X denotes an unknown state.  
Therefore, we employ the technique from \eqref{piar_eq} to evaluate global channel congestion and, similar to PB \cite{PAR09}, piggyback the global channel congestion bitmap within the group, enabling source routers to accurately perceive intra-group global channel congestion. The piggybacking mechanism itself is inherited directly from PB: the congestion bitmap rides on regular intra-group traffic and, absent data packets, can be carried by periodic control messages between adjacent routers---an in-hardware congestion-feedback capability the Slingshot~400 switch ships~\cite{ss400}.

\begin{algorithm}[t] 
\footnotesize
  \caption{DIAR Routing Algorithm} 
  
  \label{recn_algo}
  \begin{algorithmic}[1]
    \Statex
    \hspace{-1em}
    \begin{tabular}{l@{:~}l}
      $q_m (q_{nm})$  & queue occupancy for minimal (non-minimal) path\\  
      $C_m (C_{nm})$  & congestion status for minimal (non-minimal) global channel\\
      $l$ & one-way latency of the local channel (in cycles)\\
    \end{tabular}
    \Statex \textit{// Global Adaptive Routing}
    \If{$!C_m \ \ or \ \ C_{nm}$}
        \State \text{Route global minimally}
    \Else
    \State \text{Route global non-minimally}
    \EndIf
    \Statex \textit{// Local Incremental Adaptive Routing}
    \If{global minimal \textit{and} next minimal hop is a local channel}
    \If{$q_m \leq 2l \ \ or \ \ q_m \leq 2\times q_{nm}$}
        \State \text{Route local minimally}
    \Else
    \State \text{Route local non-minimally}
    \EndIf
    \EndIf
    
  \end{algorithmic}
\end{algorithm}

\subsection{Decoupled, Hierarchical Two-Stage Adaptive Routing}
\label{sec43}
Unlike source routing algorithms, which simply classify routing strategies into global minimal routing and global non-minimal routing, DIAR finely decouples them into hierarchical global and local incremental adaptive routing stages. Lines 1--5 represent the global adaptive routing stage, which is executed exclusively when a packet arrives at the source router. Here, \( C_m \) (\( C_{nm} \)) denotes the congestion status of the global channel on the minimal (non-minimal) path. A value of true indicates that the channel is congested; otherwise, it is not congested. Both flags are read directly from the piggybacked intra-group bitmap (Section~\ref{sec42}): $C_m$ indexes the bit of the minimal path's global channel, and $C_{nm}$ that of the global channel toward the randomly pre-selected Valiant intermediate group. When the global channel of the minimal path is not congested, or the global channel of the non-minimal path is congested, global minimal routing should be selected; otherwise, global non-minimal routing should be applied. 
Lines 6--12 represent the local incremental stage, which executes for global minimal packets (line 6, Rule (3)), exactly twice per packet: at the source router and at the entry router of the destination group; the guard also skips the stage when the minimal next hop is already the global channel or the destination. Here, $q_m$ ($q_{nm}$) denote the downstream queue occupancies of the minimal output port and of the detour candidate port---a randomly selected router within the group. Because the VC of an arriving packet encodes whether a detour has already been taken (e.g., arriving on $L_{vc2}$ implies a completed detour), and because, in a fully connected group, a detour hop always terminates at a router directly connected to the group's exit or to the destination, at most one detour per group is enforced structurally---the six-hop bound requires no hop counters or additional header bits. When both global channels are reported congested, the packet is routed minimally, since a Valiant detour would offer no congestion advantage while consuming twice the global bandwidth.
Here, \( l \) denotes the one-way latency of the local channel; assuming single-cycle routers, \( 2l \) corresponds to the round-trip latency of the channel. Under steady, benign transmission, the queue occupancy of a channel (i.e., the credits consumed on that channel) never exceeds its round-trip latency.
Local non-minimal routing should only be performed when necessary: to avoid the adverse impact of transient traffic and phantom congestion \cite{par2w15}, packets are routed local minimally when the queue length of the minimal path is below the round-trip latency of the channel or does not exceed twice the queue length of the non-minimal path; only otherwise is local non-minimal routing performed.
Following prior works \cite{DGB22, piar25, par2w15}, this paper adopts the single-cycle router model by default; real routers incur multi-cycle delays, compensated by adding the extra latency to \( 2l \) \cite{par2w15}.

Compared to other Dragonfly routing algorithms \cite{DGB22, Q_ada21, piar25}, DIAR is remarkably simple, which is precisely the charm of incremental routing. Since congestion can occur at any hop along the minimal path, source routing algorithms require the source router to perceive distant congestion as much as possible, which has led to various heuristics for estimating remote congestion, such as machine learning-based approaches and end-to-end latency piggybacking \cite{Q_ada21, DGB22}. These methods significantly increase the design complexity of the routing algorithm. In contrast, DIAR decomposes routing into two sub-stages: for the global routing stage, DIAR only needs to consider whether the global channel is congested, regardless of local channel conditions. Furthermore, routers within the source group do not need to care whether the local channels in the destination group are congested, as the decision to employ a local detouring strategy is delegated to the routers in the destination group. Ultimately, through the combination of the two-stage routing strategy, DIAR can fully exploit all almost-minimal paths in the network. We will demonstrate the superior performance of DIAR in the next section.

\section{PERFORMANCE}
\begin{table}[t] 
\footnotesize
  \caption{Simulation parameters}
  \label{simulator}
  \centering
  \begin{tabular}{|l|l|}
    \hline
    Parameter & Value \\
    \hline
    Topology           & dfly(4,8,4,33)\\
    \hline
    Number of nodes    & 1056\\
    \hline
    Switch Speedup & 2.0\\
    \hline
    Number of VCs &2 (MIN), 5 ($\rm{PAR_{PH}}$), 4 (others)\\
    \hline
    VC Buffer size (flits) & 32 (local), 256 (global) \\
    \hline
    Channel latency (cycles)  & 10 (local), 100 (global) \\
    \hline
    Routing algorithms& MIN \cite{Dragonfly08}, VAL \cite{val82}, PIAR \cite{piar25}, UGAL \cite{ugal05}\\
    &  $\rm{PAR_{PH}}$ \cite{PAR09, par2w15}, UGAL-LE \cite{ugal-le24} \\
    & DGB \cite{DGB22}, DIAR (This paper)\\
    \hline
    Hyperparameters &  DIAR (T = 25), DGB ($\alpha=0.1$, T = 50 \cite{DGB22}) \\
    &PIAR ($Th_l:10$, $Th_{g1}:40$, $Th_{g2}:10$ \cite{piar25})\\
     &  UGAL-LE (C = 5 \cite{ugal-le24}) \\
    \hline
    History window size &  DIAR (200 for global links)\\
   (cycles) &  UGAL-LE (500 \cite{ugal-le24}) \\
     &  $\rm{PAR_{PH}}$, DGB and PIAR \\
     & (20/200 for local/global links \cite{piar25, DGB22, par2w15}) \\
     
\hline
\end{tabular}
\end{table}

\subsection{Methodology}

The synthetic workload evaluations were conducted using BookSim 2.0 \cite{booksim13}, a cycle-accurate network simulator widely adopted in prior works. We utilized the identical experimental configurations as previous studies \cite{DGB22, par2w15, piar25}, summarized in Table \ref{simulator}. Following many prior works \cite{Q_ada21,T_ugal19, ics20_global_connect, piar25, DGB22}, an input-queued, single-cycle router model with sufficient internal speedup is adopted to eliminate router bottlenecks, each simulation begins with an unrecorded warm-up phase to reach steady state, single-flit packets are employed to exclude flow-control effects, and the bias parameter ($b$ in \eqref{ugal_eq}) is set to 0 for UGAL and UGAL-LE---the only evaluated algorithms that use it---so that the routing mechanism favors neither minimal nor non-minimal routing. We report throughput and average packet latency, and consider the network saturated once the latter exceeds 400 cycles; saturation is also directly visible in the latency--throughput curves as a sharp latency upturn.

Notably, all compared algorithms are taken from the open-source implementations \cite{dgb_booksim2, piar_source_code} of DGB \cite{DGB22} and PIAR \cite{piar25}---both state-of-the-art on Dragonfly---which cover every baseline, ensuring identical configurations and faithful reproduction.
\begin{figure*}[t]
  \centering

  \begin{subfigure}[t]{0.24\linewidth}
    \centering
    \includegraphics[width=\linewidth]{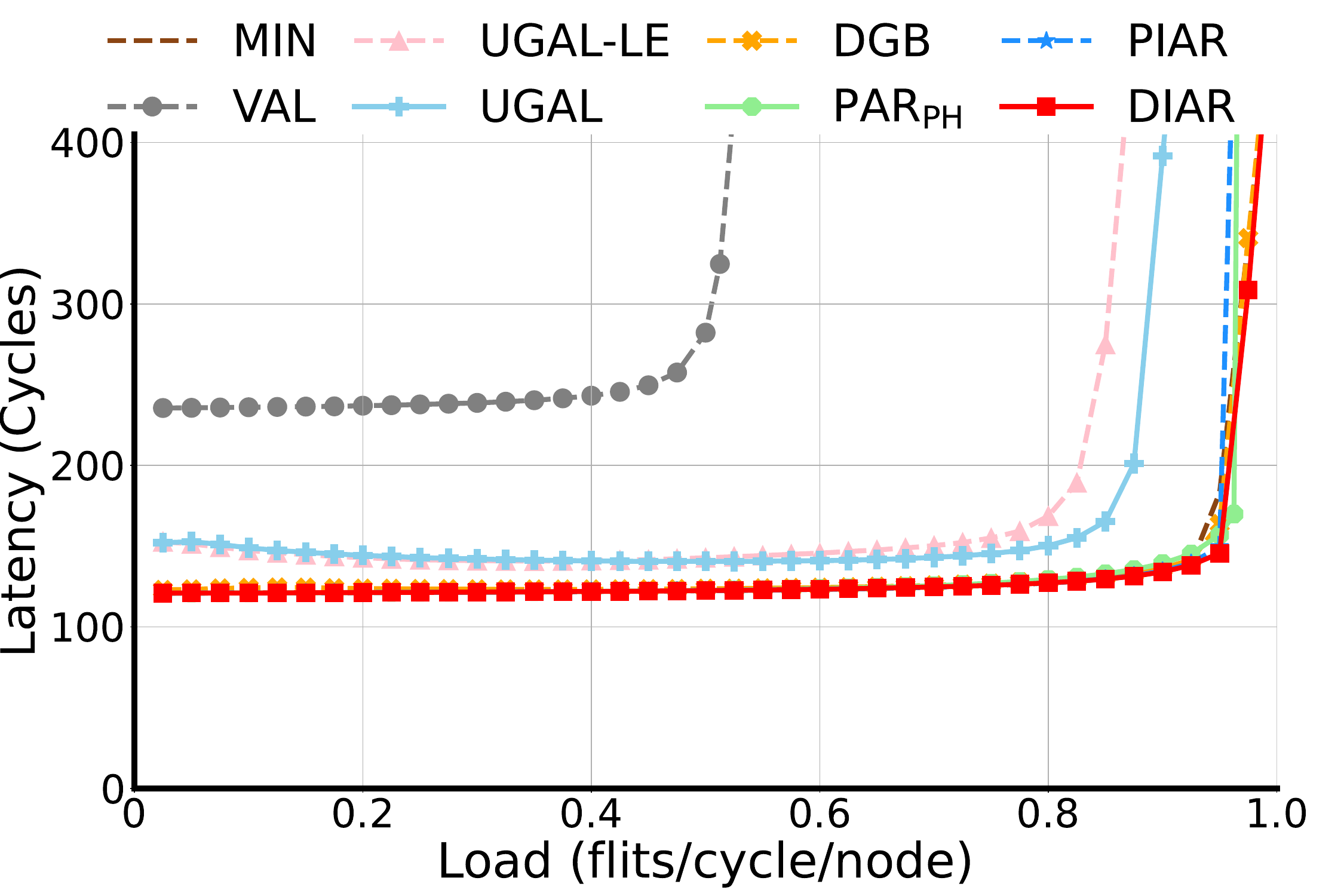}
    \caption{Uniform Random}
    \label{ur_performance}
  \end{subfigure}
  \hfill
  \begin{subfigure}[t]{0.24\linewidth}
    \centering
    \includegraphics[width=\linewidth]{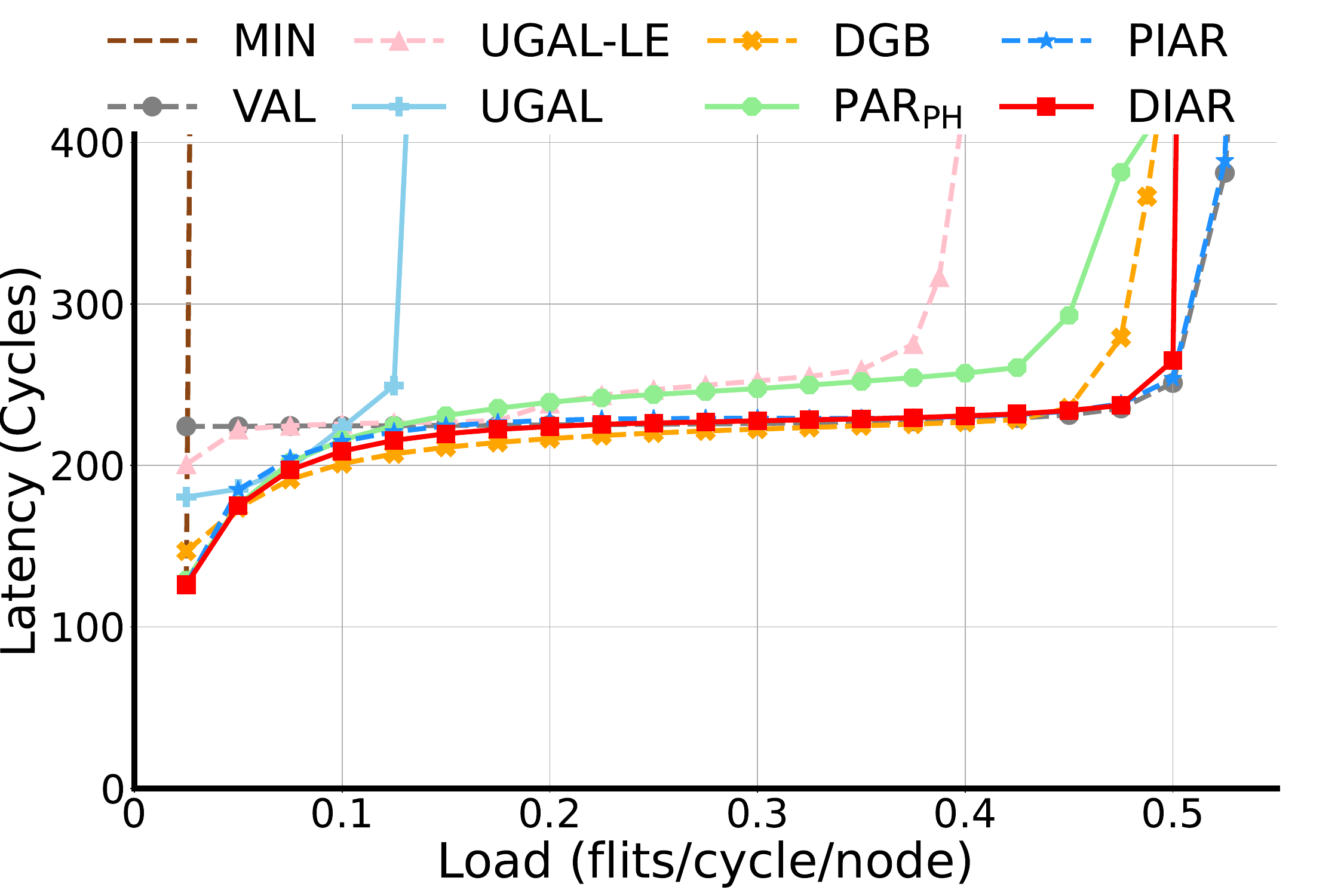}
    \caption{Adversarial}
    \label{adv_performance}
  \end{subfigure}
  \hfill
  \begin{subfigure}[t]{0.24\linewidth}
    \centering
    \includegraphics[width=\linewidth]{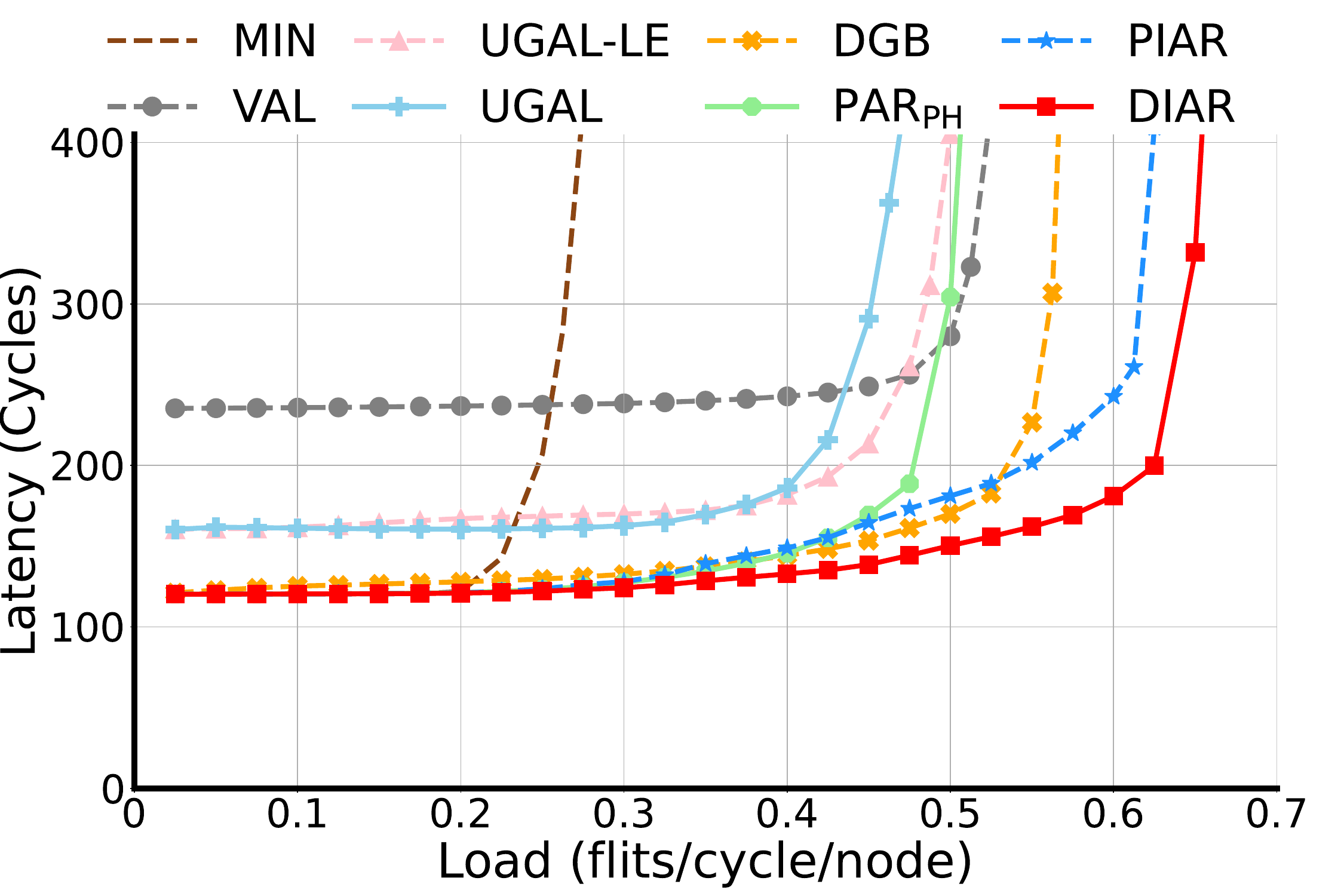}
    \caption{Random Permutation}
    \label{perm_performance}
  \end{subfigure}
  \hfill
  \begin{subfigure}[t]{0.24\linewidth}
    \centering
    \includegraphics[width=\linewidth]{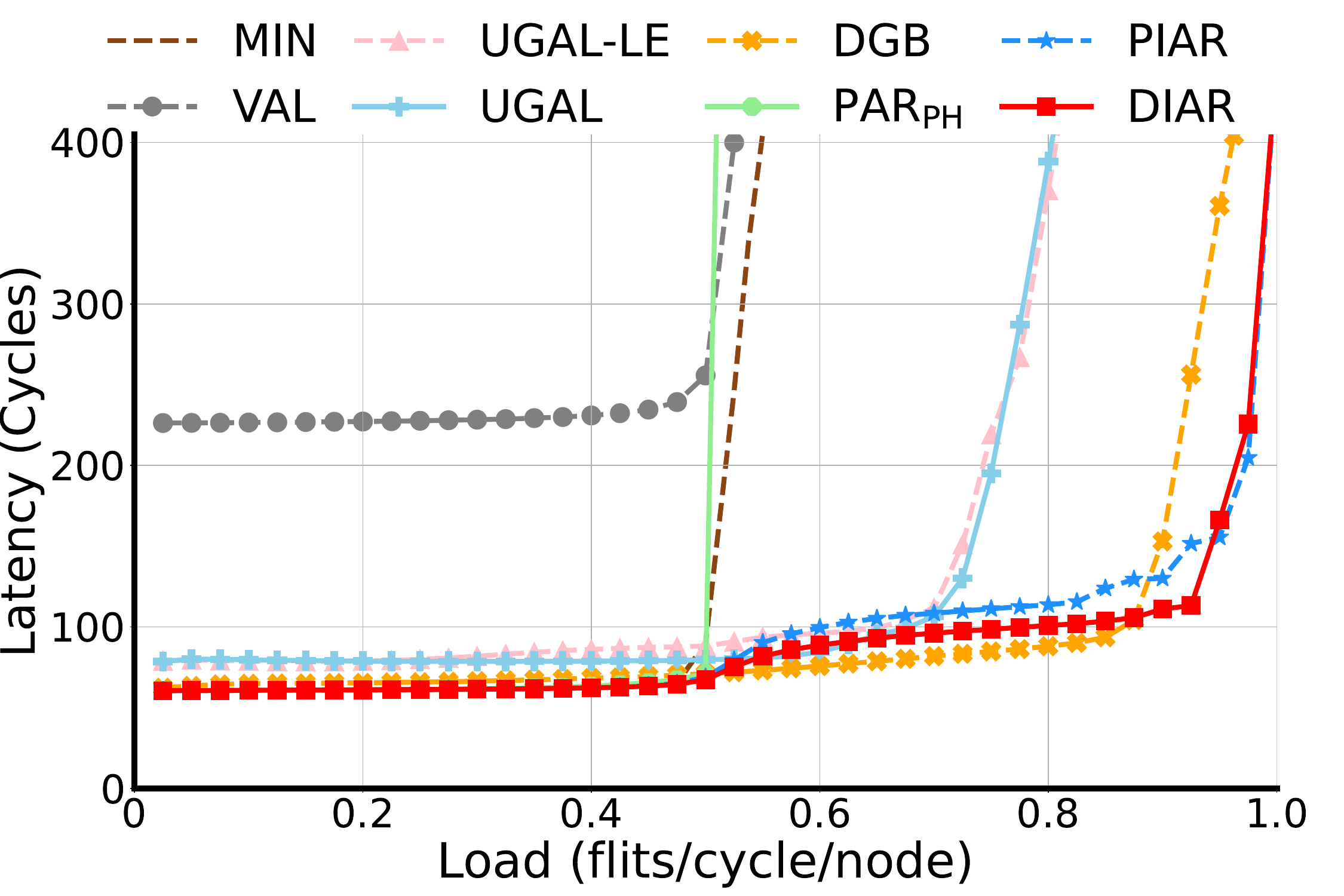}
    \caption{$\rm{UR_{half}}$}
    \label{figHalf}
  \end{subfigure}
  \caption{Latency-throughput comparison for various traffic patterns on a 1,056-node Dragonfly network (dfly(4,8,4,33)).}
  \label{mixed_performance}
\end{figure*}

For the HPC workload evaluation, similar to \cite{DGB22, piar25}, we integrate the SST simulator \cite{sst11} with BookSim, where traffic is generated by SST and injected on-the-fly into BookSim. The link bandwidth is set to 200~Gbps. The configurations are identical to those in Table~\ref{simulator}.

\subsection{Synthetic Traffic Patterns}
\textbf{1) Uniform Random (UR):}  In the UR traffic pattern, the destinations of packets are randomly selected from the network. Ideally, all traffic should be routed minimally. Figure \ref{ur_performance} illustrates the performance of various routing algorithms under the UR traffic pattern. DIAR, PIAR, and DGB achieve good performance. UGAL and UGAL-LE experience performance degradation due to phantom congestion \cite{par2w15}, which causes part of the traffic to be routed along non-minimal paths. UGAL, UGAL-LE, and $\rm{PAR_{PH}}$ achieve throughputs of 0.9, 0.8625, and 0.9625, respectively, whereas DIAR achieves a throughput of 0.975, improving performance over them by 8.3\%, 13.0\%, and 1.3\%, respectively.

\textbf{2) Adversarial (ADV):} In the ADV traffic pattern, the nodes in group $i$ send all their traffic to random nodes in group $i+1$, thereby stressing inter-group channels. Ideally, the vast majority of traffic should be routed via non-minimal paths. Figure \ref{adv_performance} presents the performance of the evaluated routing algorithms under the ADV pattern. The minimal routing algorithm achieves extremely low throughput, and UGAL also suffers from low throughput due to the indirect routing issue \cite{Dragonfly08, par2w15, DGB22}. By leveraging local sliding windows to better sense the congestion of remote global channels, UGAL-LE improves upon UGAL's performance, but the improvement is limited; $\rm{PAR_{PH}}$, which can additionally correct routing decisions en route, performs considerably better. DGB, PIAR, and DIAR all achieve good performance. Specifically, DIAR improves throughput over UGAL-LE and $\rm{PAR_{PH}}$ by 29.0\% and 5.3\%, respectively; moreover, whereas UGAL achieves a throughput of only 0.125, DIAR reaches 0.5---a 4$\times$ improvement.

\textbf{3) Random Permutation (PERM):} In the random permutation pattern, each node sends traffic exclusively to a single, unique destination node. PERM is a complex traffic pattern that simultaneously creates ADV-like scenarios stressing inter-group channels and scenarios stressing local channels as illustrated in Figure~\ref{LocalCongestion}(c)--(f). Ideally, routing should balance load with the fewest extra hops; as shown in Section~\ref{congestion_model}, about one-third of the contended paths here can be resolved by almost-minimal paths. However, as shown in Figure \ref{perm_performance}, all algorithms except DIAR fail to exploit the almost-minimal paths in the network, which wastes global bandwidth and results in higher latency and lower throughput.
UGAL, UGAL-LE, and \(\rm{PAR_{PH}}\) achieve relatively low throughput because they cannot accurately sense the congestion of remote global channels. DGB and PIAR sense global congestion more accurately and achieve better global load balancing, but their throughput is likewise limited by their inability to exploit almost-minimal paths. PIAR outperforms DGB because it exploits global non-minimal paths with fewer hops. In contrast, DIAR achieves global load balancing while fully exploiting almost-minimal paths, attaining the lowest latency and the highest throughput of 0.65. Compared to DGB and PIAR, DIAR improves throughput by 15.6\% and 6.1\%, respectively, and by 40.5\%, 33.3\%, and 30.0\% over UGAL, UGAL-LE, and $\rm{PAR_{PH}}$, respectively.

\textbf{4) $\rm{UR_{half}}$ \cite{DGB22}:} In the $\rm{UR_{half}}$ pattern, the network is linearly partitioned into two equal halves occupied by Job 1 and Job 2, respectively. Job 1 is communication-intensive and generates UR traffic internally, whereas Job 2 is computation-intensive and generates no network traffic. Under this pattern, the minimal paths within Job 1's region can utilize only half of the global channels, effectively forming a 2:1 oversubscribed Dragonfly subnetwork.
As shown in Figure~\ref{figHalf}, MIN performs poorly because it can exploit only half of the global channels within Job 1. $\rm{PAR_{PH}}$ performs very similarly to MIN: since $\rm{PAR_{PH}}$ averages the queue lengths across all non-minimal ports, it overestimates the congestion of the global channels leading to Job 2 and thus routes excessive traffic minimally. UGAL-LE observes the ADV traffic sent by other routers within the group through its local sliding windows; however, the 2:1 oversubscription is insufficient to reach UGAL-LE's congestion threshold, causing it to degenerate into UGAL with a throughput of 0.8. DGB, PIAR, and DIAR, which piggyback global channel congestion information within the group, possess finer-grained awareness of global channels and achieve better throughput. Overall, DIAR improves throughput over DGB, UGAL-LE, and $\rm{PAR_{PH}}$ by 2.6\%, 21.9\%, and 95.0\%, respectively.

\begin{figure}[t]
  \centering
  \includegraphics[width=0.98\linewidth]{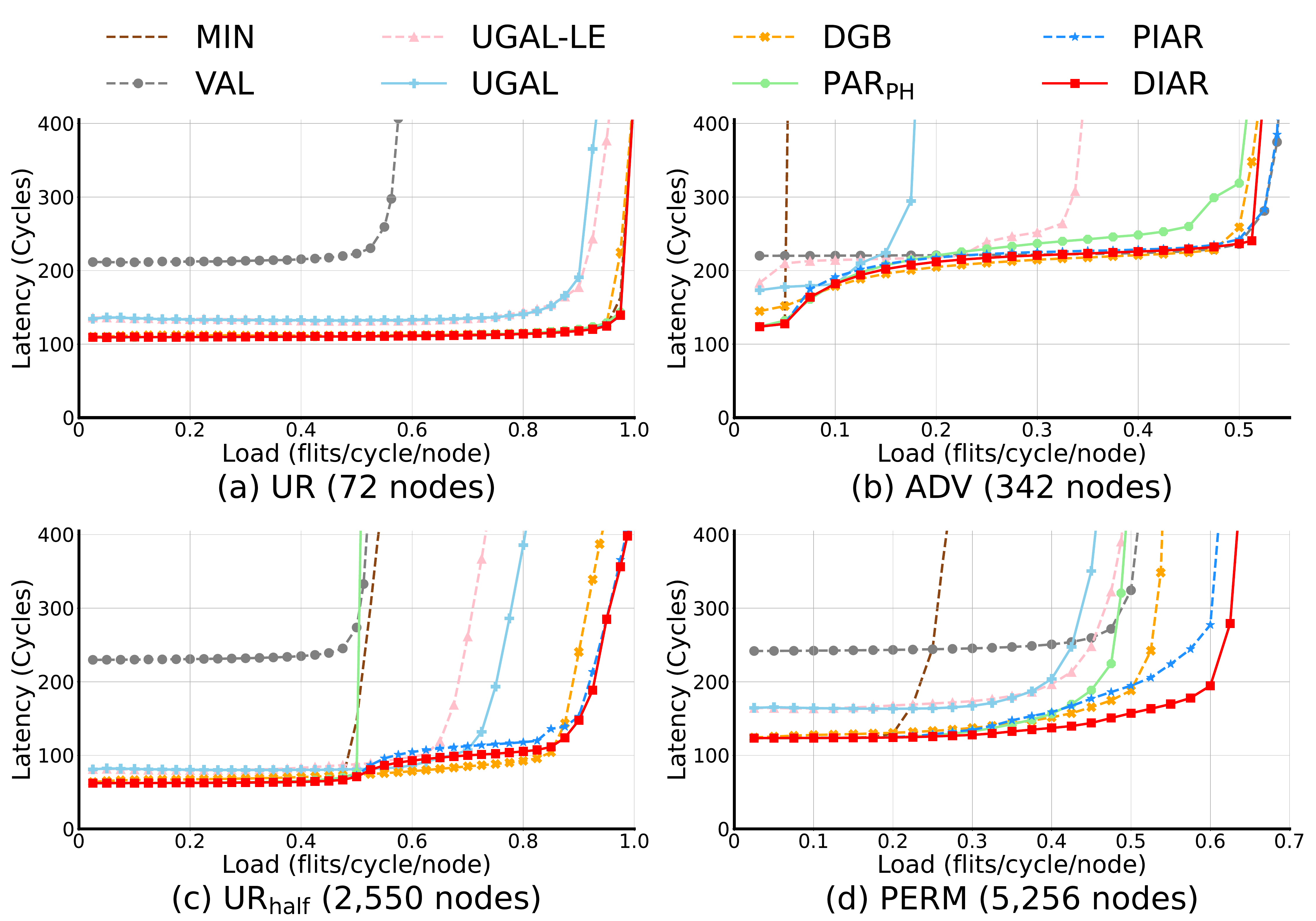}
  \caption{Performance on Dragonfly networks of different scales, one representative traffic pattern per scale; hyperparameters unchanged (Table \ref{simulator}).}
  \label{scalability_performance}
\end{figure}

\textbf{5) Scalability:} We further evaluate four balanced Dragonfly configurations ($p=h$, $a=2h$, $g=ah+1$, $p\in\{2,3,5,6\}$, 72--5,256 nodes)---two orders of magnitude in scale---each with one representative traffic pattern. As shown in Figure \ref{scalability_performance}, the relative ranking of the algorithms is essentially unchanged, and DIAR delivers the highest (or tied-highest) throughput except under ADV, where it remains on par with DGB; its saturation throughput deviates from the 1,056-node counterpart by at most 3.8\% with $T$ fixed at 25.

\begin{figure*}[!t]
  \centering
  \includegraphics[width=0.98\linewidth]{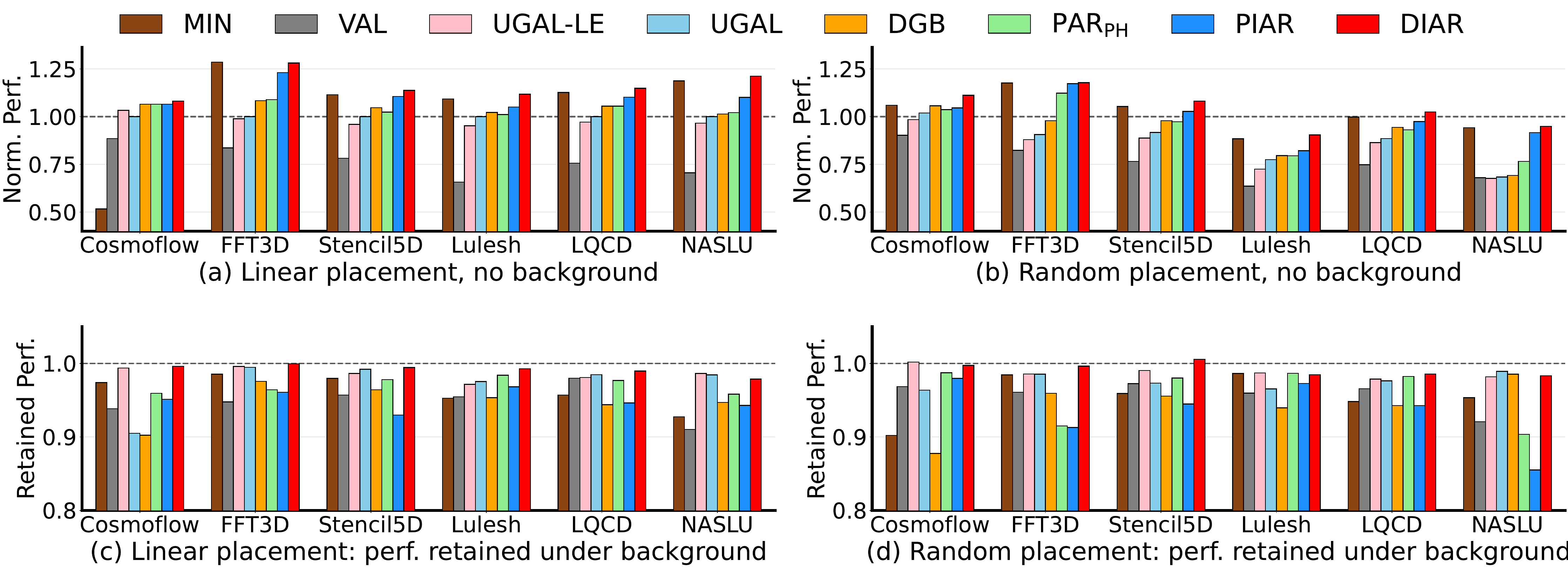}
  \caption{Performance on six HPC workloads (512-rank job, dfly(4,8,4,33)). (a)--(b): performance (1/completion time) normalized to UGAL under linear placement without background traffic. (c)--(d): fraction of performance retained when the remaining 544 nodes run an intra-group pairwise all-to-all background job. Y-axes start at 0.4 in (a)--(b) and 0.8 in (c)--(d).}
  \label{app_performance}
\end{figure*}

\subsection{HPC Workloads}
\label{sec5C}

We evaluate the six HPC workloads from \cite{qada22}, covering the major communication patterns of production applications: Cosmoflow performs tree-based Allreduce; FFT3D performs all-to-all exchanges within row and column communicators; Stencil5D performs a five-dimensional halo exchange with ten neighbors per rank; LULESH combines 26-neighbor nearest-neighbor exchange with sweep and collective phases; LQCD performs a four-dimensional lattice boundary exchange; and NAS-LU performs a pipelined 2-D wavefront sweep. Each workload runs as a 512-rank job communicating roughly 100~KB per rank, using the configurations in Table~\ref{simulator}.

The 512 compute nodes are randomly sampled from the 1,056-node network, and logical ranks are mapped onto these physical nodes either \emph{linearly} (rank order follows node order) or \emph{randomly}. In addition, we evaluate a background noise scenario: the remaining 544 nodes continuously run an intra-group pairwise all-to-all exchange. Since communication locality is pervasive in these workloads---as the placement results below confirm---an intra-group job is also the noise a co-runner most likely injects. This background traffic never uses global channels, but it saturates a subset of the local channels within every group---precisely the congestion scenarios illustrated in Figure~\ref{LocalCongestion}(a)--(b). Figure~\ref{app_performance}(a)--(b) reports each algorithm's performance without background traffic, and (c)--(d) the fraction of that performance retained under the background job.

Without background noise, DIAR achieves the best performance among all the adaptive routing algorithms, outperforming PIAR, DGB, UGAL-LE, $\rm{PAR_{PH}}$, and UGAL by up to 10.1\%, 37.2\%, 40.3\%, 24.0\%, and 38.9\%, respectively. Moreover, linear placement generally outperforms random placement, as it preserves this locality. MIN performs well in most cases, but degrades severely on Cosmoflow under linear placement ($2.1\times$ slower than DIAR): the reduction tree concentrates each level of communication onto a few pairs of groups, saturating the single global channel between every such pair and effectively producing ADV-like traffic that minimal routing cannot avoid.

Under background noise, DIAR again delivers the best performance and is barely affected: by detouring around the congested local channels via almost-minimal paths, it loses at most 2.1\% of its performance. In contrast, PIAR, DGB, $\rm{PAR_{PH}}$, UGAL, and UGAL-LE lose up to 14.5\%, 12.2\%, 9.6\%, 9.5\%, and 2.8\% of their performance, respectively. Compared with the noise-free case, DIAR's maximum advantages over PIAR, DGB, UGAL-LE, $\rm{PAR_{PH}}$, and UGAL become 19.2\%, 37.0\%, 40.5\%, 34.9\%, and 38.1\%, respectively.

\begin{figure}[!t]
  \centering
  \includegraphics[width=0.98\linewidth]{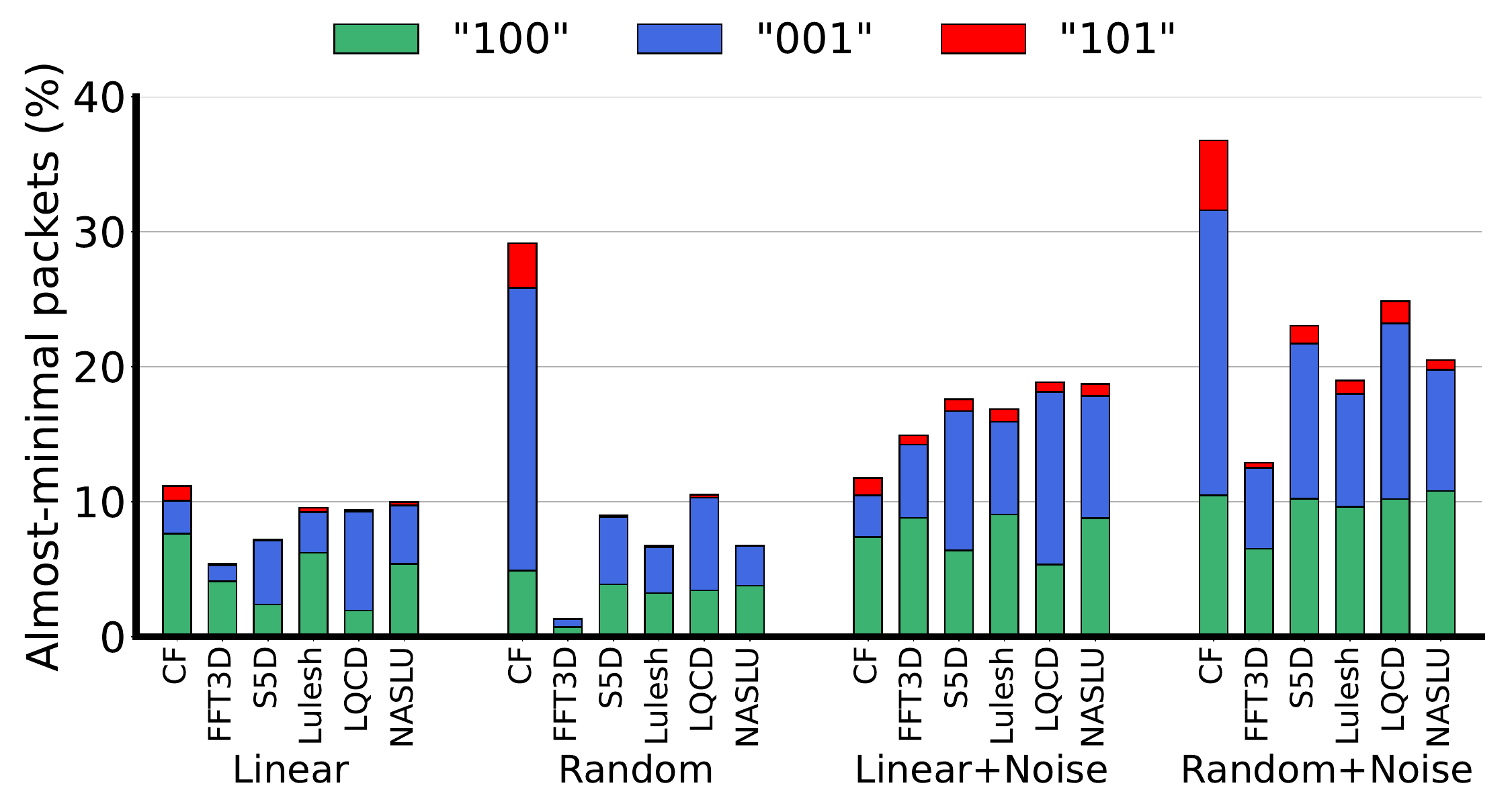}
  \caption{Fraction of DIAR packets that traverse each
  type of almost-minimal path on the six HPC workloads, under both
  placements, with and without background noise.}
  \label{am_usage}
\end{figure}

Figure~\ref{am_usage} corroborates this mechanism, reporting the
measured fraction of DIAR packets traversing each class of
almost-minimal path~\eqref{am_seqs}. Even without background noise,
1.3\%--29.2\% of packets take almost-minimal detours---most notably on
Cosmoflow under random placement, where destination-group detours
(``001'') alone reach 21\% as packets sidestep the reduction-tree
incast. Once the background job congests the local channels, the usage
roughly doubles across the board (e.g., 5.4\%$\rightarrow$14.9\% on
FFT3D and 9.4\%$\rightarrow$18.8\% on LQCD under linear placement),
while double detours (``101'') stay rare ($\leq$5.2\%); DIAR's noise
resilience thus comes from these local side-steps rather than from
global non-minimal rerouting.

Overall, regardless of the placement strategy and the presence of background traffic, DIAR always achieves the best performance among the adaptive routing algorithms---leading the strongest realizable baseline in each of the 24 cells by 1.6\%--34.9\% (13.2\% on average)---and retains the most performance under the evaluated intra-group background traffic.

\section{COST AND SENSITIVITY ANALYSIS}

\begin{figure*}[!t]
  \centering

  \begin{subfigure}[t]{0.32\linewidth}
    \centering
    \includegraphics[width=\linewidth]{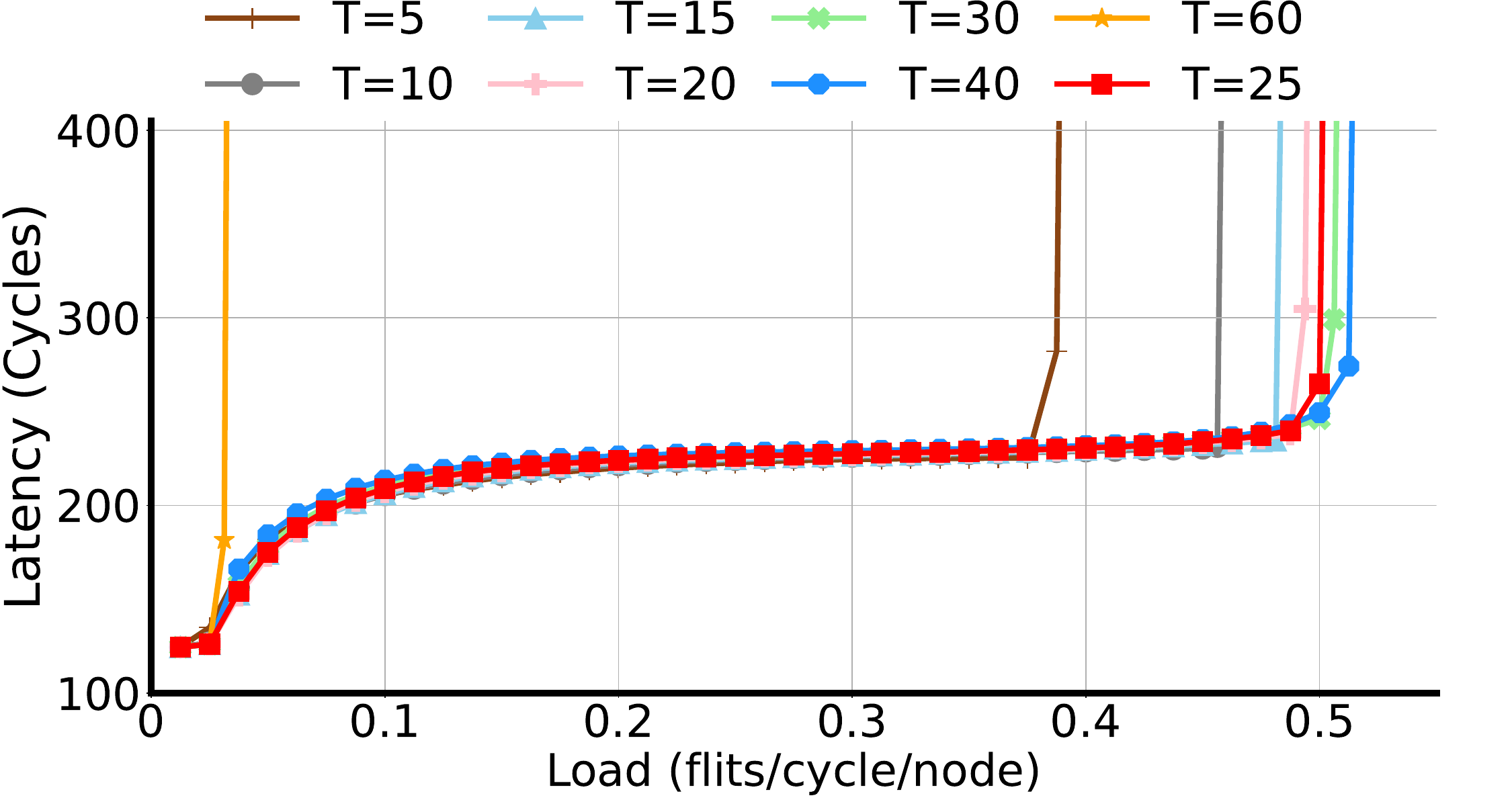}
    \caption{Adversarial}
    \label{fig:mixed}
  \end{subfigure}
  \hfill
  \begin{subfigure}[t]{0.32\linewidth}
    \centering
    \includegraphics[width=\linewidth]{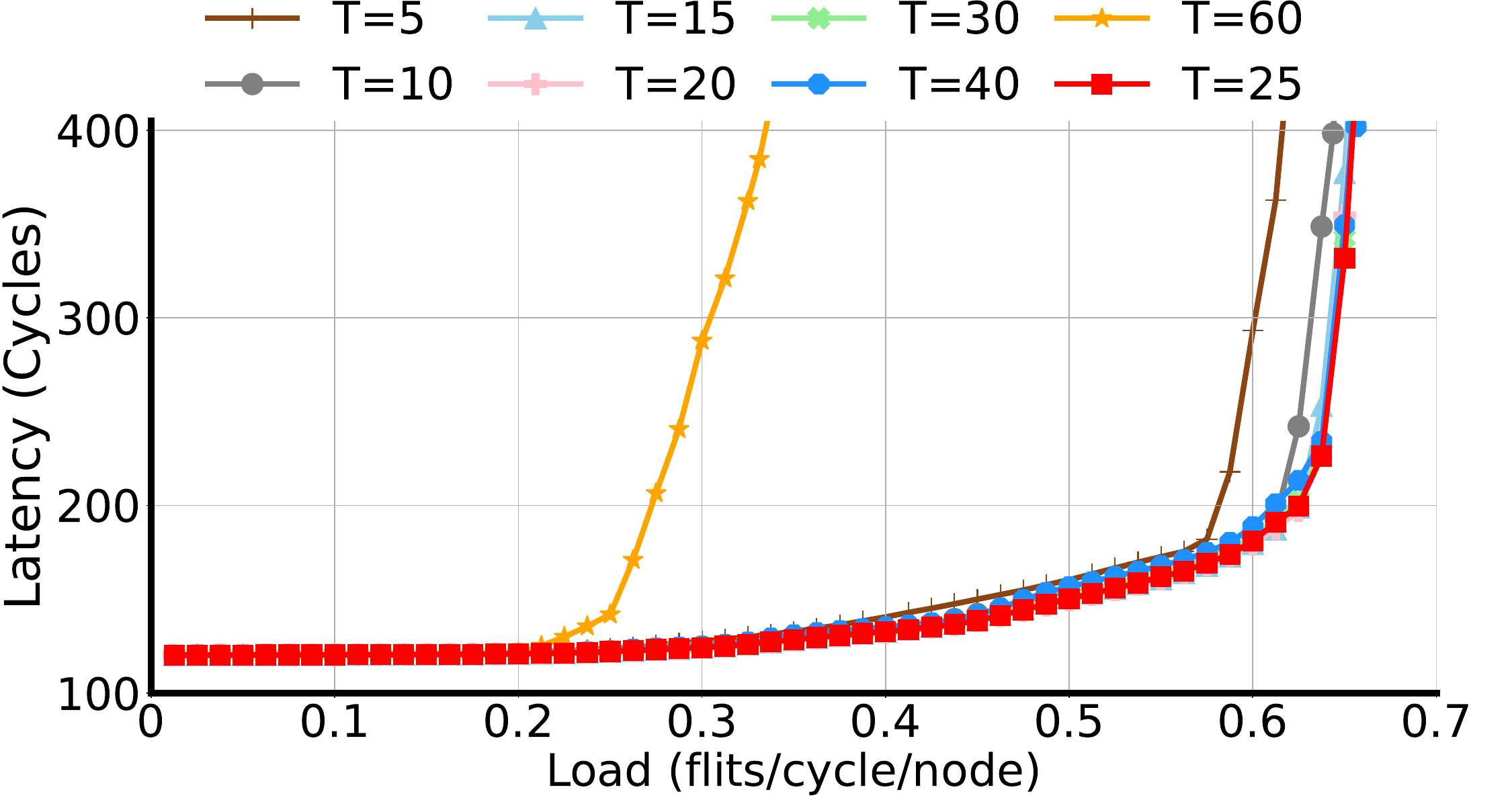}
    \caption{Random Permutation}
    \label{fig:ur}
  \end{subfigure}
  \hfill
  \begin{subfigure}[t]{0.32\linewidth}
    \centering
    \includegraphics[width=\linewidth]{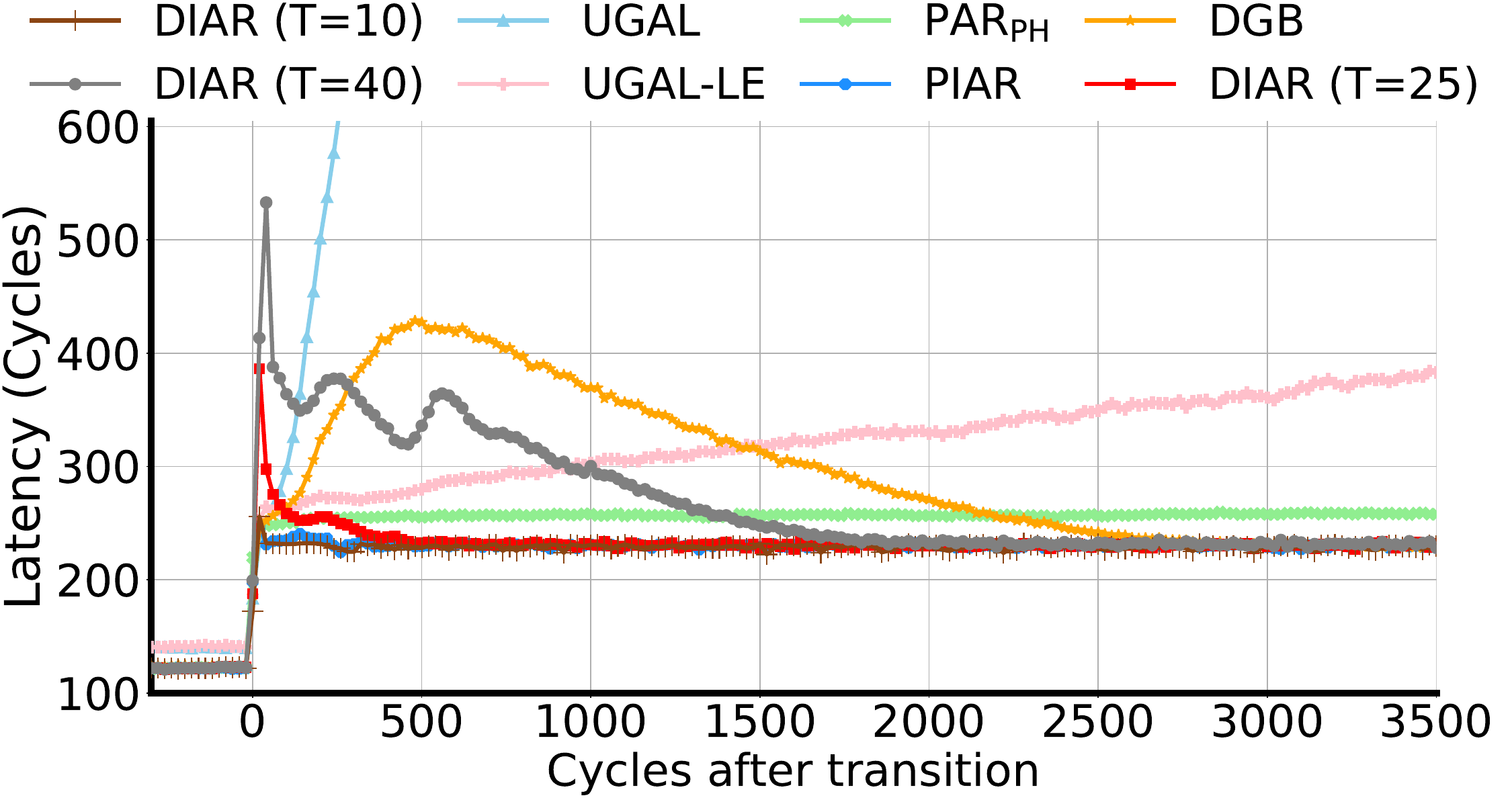}
    \caption{Dynamic traffic}
    \label{transient}
  \end{subfigure}

  \caption{Sensitivity of DIAR. (a)--(b): latency-throughput under different congestion thresholds. (c): latency when traffic changes from UR to ADV at load 0.4.}
  \label{sensitivity}
\end{figure*}

\begin{table}[t]
\footnotesize
  \caption{Implementation overheads of various routing algorithms. The buffer overhead depends on the router radix (\(k\)) and the sliding window length (\(L\)). Symbols \gou, \bangou, and \cha ~ represent low, medium, and high overhead, respectively.}
  \label{cost_comp}
  \begin{tabular}{|c|c|c|c|c|}
    \hline
    Algorithm&Buffer Overhead & Buffer & Logic & Bandwidth\\
    \hline
    UGAL \cite{ugal05}& $\mathcal{O}(1)$ & \gou & \gou & \gou \\
    \hline
    $\rm{PAR_{PH}}$ \cite{par2w15}& $\mathcal{O}(kL)$ & \gou & \bangou &\gou\\
    \hline
    UGAL-LE \cite{ugal-le24} & $\mathcal{O}(k^3L)$ & \cha & \gou & \gou \\
    \hline
    DGB \cite{DGB22} & $\mathcal{O}(k^4+kL)$ &\cha&\cha&\cha\\
    \hline
    PIAR \cite{piar25} & $\mathcal{O}(k^2+kL)$ &\bangou &\cha&\gou\\
    \hline
    DIAR& $\mathcal{O}(kL)$ & \gou & \gou & \gou\\
  \hline
\end{tabular}
\end{table}

\label{Overhead}
\subsection{Implementation Cost}

Although numerous adaptive routing algorithms have been proposed for Dragonfly networks, as shown in Table~\ref{increment_table}, only variants of UGAL and PAR have been implemented on router chips \cite{par2w15}---many high-performing algorithms incur prohibitive hardware overhead. In contrast, DIAR simultaneously achieves high performance and lightweight hardware overhead, as shown in Table~\ref{cost_comp}.

For buffer overhead, let \(h\) denote the total number of global ports in each router. DIAR requires only \(h\) sliding windows, and the length of each window equals the round-trip latency of the global channels.

$\rm{PAR_{PH}}$ maintains the same class of sliding windows as DIAR.\footnote{This accounts only for the sliding-window state; $\rm{PAR_{PH}}$'s additional fifth VC further enlarges the router's entire data buffer by 25\%.} UGAL-LE utilizes a larger number of sliding windows to track the destinations of intra-group non-minimal traffic, incurring a considerably higher buffer overhead, whereas DGB must maintain the congestion states of all non-minimal paths between router pairs, resulting in a prohibitive buffer overhead. PIAR stores hierarchical tables of shorter-hop paths, a moderate additional overhead. For the network evaluated in this paper with a router radix of 15 (estimated as 16 ports, rounded up to the next power of two) and 1,056 nodes, DIAR requires only 100~Bytes of buffer space, whereas UGAL-LE, DGB, and PIAR require 16.5~KB, 9.3~KB, and 204~Bytes, respectively. When considering routers with 64 ports (as in Slingshot~11) and future 128 ports, UGAL-LE and DGB will consume several to tens of megabytes of precious on-chip SRAM---a heavy burden for UGAL-LE and almost impossible to realize for DGB (Figure~\ref{buffer_cost}). At such scales, DIAR cuts buffer overhead by roughly $5\times$ versus PIAR and by up to four orders of magnitude versus UGAL-LE and DGB.

While a high buffer overhead is theoretically manageable, the true barrier lies in logic complexity. $\rm{PAR_{PH}}$ requires an integer division and one additional VC, whereas DGB demands consecutive floating-point multiplications and divisions with tight logical dependencies, which are hard to parallelize. PIAR avoids floating point but must, for every packet, sequentially traverse hundreds to thousands of candidate non-minimal paths across multiple routing tables according to \eqref{piar_eq}---up to 512 candidates on a 64-port router and 2,048 on a 128-port one. Practical router architectures instead sample only a few candidates via hashing or randomization, since traversing the entire set would stall the data path or reduce the clock frequency \cite{hyperXRouting19}. This constraint persists in the newest hardware: the Slingshot~400 switch performs ``packet-by-packet adaptive path selection using local and global load'' while forwarding 64-byte packets at line rate on 400\,Gbps ports---781 million packets per second per port~\cite{ss400}. Regardless of pipeline depth, the routing logic of every port must therefore initiate a new decision roughly every 1.3\,ns.
In contrast, implementing DIAR is far simpler: as shown in Algorithm~\ref{recn_algo}, it requires no complex computational operations, relying only on a few comparators, shifters, adders, and multiplexers---primitives the shipped Cray Aries router already employs (no floating point) \cite{cray12, crayRout17}.

For bandwidth overhead, PIAR, DGB, and DIAR all need to piggyback congestion information between routers, leading to a bandwidth loss of \(\frac{P}{D}\), where \(P\) and \(D\) denote the piggybacked-information size and the packet length \cite{PAR09}. Both PIAR and DIAR piggyback congestion bitmaps of \(h\) global channels within a group. With 256-Byte packets, PIAR and DIAR incur only 0.2\% overhead in the evaluated network. By contrast, DGB piggybacks a quintuple of global congestion information (destination router ID, end-to-end latency, etc.~\cite{DGB22, dgb_booksim2}); at 10 bytes, its overhead reaches 3.9\%.

Overall, when performance is weighed against implementation cost---an aspect largely overlooked by prior works---DIAR emerges as a highly cost-effective solution; RTL synthesis of the routing decision logic is left for future work.

\label{Sensitivity}
\subsection{Sensitivity of DIAR}
The congestion threshold \(T\) in \eqref{piar_eq} is DIAR's only tunable hyperparameter---the local-stage constants of Algorithm~\ref{recn_algo} are fixed by the round-trip argument in Section~\ref{sec43}---and is set to 25 by default. Figure \ref{sensitivity} shows DIAR under ADV and PERM for different \(T\): a small \(T\) lets instantaneous fluctuations be misread as global-channel congestion, limiting performance at high injection, whereas an overly large \(T\) (e.g., 60) delays congestion detection and hence the piggybacked information; within 20--40 the variations are negligible. The HPC-workload sensitivity likewise resembles PERM, varying no more than 3.2\% within the same range.

The impact of \(T\) is best seen in the transient load scenario of Figure~\ref{transient}. UGAL and UGAL-LE cannot withstand severe transient load variations: UGAL's latency diverges and UGAL-LE's plateaus near the saturation threshold. DGB and DIAR (\(T=40\)) converge slowly (${\sim}$2800 and ${\sim}$2300 cycles). \(\rm{PAR_{PH}}\), PIAR, and DIAR (\(T=10\)) converge smoothly within 40 cycles. DIAR (\(T=25\)) is slightly slower (about 500 cycles), trading responsiveness for steady-state performance.

\section{CONCLUSION}

We rethink adaptive routing on Dragonfly around two overlooked issues: the performance--realizability dilemma and the unexploited almost-minimal paths. DIAR decouples global path selection at the source from incremental local detours; a six-hop bound yields deadlock freedom with the standard four VCs; the routing logic requires only a few comparators, shifters, adders, and multiplexers, and buffer overhead is cut by up to four orders of magnitude. DIAR matches or surpasses the ideal DGB and PIAR, outperforms every realizable algorithm, and best resists local-channel noise. Extending almost-minimal exploitation to Dragonfly+ is left for future work.

\bibliography{Reference}
\end{document}